\documentclass{article}
\usepackage{natbib}
\usepackage{setspace}

\usepackage{authblk}
\usepackage[utf8]{inputenc}
\usepackage{amsmath}
\usepackage{graphicx}
\usepackage{comment}
\usepackage{enumerate}
\usepackage{siunitx}
\usepackage{booktabs}
\usepackage{numprint}
\usepackage{makecell}  
\NewExpandableDocumentCommand\mcc{O{1}m}{\multicolumn{#1}{c|}{\makecell{#2}}}
\usepackage{mathtools}
\usepackage{amssymb}
\usepackage{comment}
\newcommand{\RNum}[1]{\uppercase\expandafter{\romannumeral #1\relax}}
\usepackage{float}
\usepackage[a4paper, total={7in, 10in}]{geometry}

\usepackage{algpseudocode}
\usepackage{algorithm}
\usepackage{lipsum}
\usepackage{stix}
\DeclareMathAlphabet\mathbfcal{LS2}{stixcal}{b}{n}
\usepackage{amsmath}
\usepackage{amsthm}
\usepackage{enumitem}
\usepackage{url}
\usepackage[figuresright]{rotating}

\title{Integrating tumor burden with survival outcome for treatment effect evaluation in oncology trials}

\author[1]{Saurabh Bhandari}
\author[1]{Michael J. Daniels}
\author[2]{Chenguang Wang}
\affil[1]{Department of Statistics, University of Florida}
\affil[2]{Regeneron Pharmaceuticals}

\date{\today}

\begin{document}

\maketitle

\abstract{In early-phase cancer clinical trials, the limited availability of data presents significant challenges in developing a framework to efficiently quantify treatment effectiveness. To address this, we propose a novel utility-based Bayesian approach for assessing treatment effects in these trials, where data scarcity is a major concern. Our approach synthesizes tumor burden, a key biomarker for evaluating patient response to oncology treatments, and survival outcome, a widely used endpoint for assessing clinical benefits, by jointly modeling longitudinal and survival data. The proposed method, along with its novel estimand, aims to efficiently capture signals of treatment efficacy in early-phase studies and holds potential for development as an endpoint in Phase 3 confirmatory studies. We conduct simulations to investigate the frequentist characteristics of the proposed estimand in a simple setting, which demonstrate relatively controlled Type I error rates when testing the treatment effect on outcomes.}

\renewcommand\thefootnote{}

\renewcommand\thefootnote{\fnsymbol{footnote}}
\setcounter{footnote}{1}

\section{Introduction}\label{sec1}

In cancer clinical studies, tumor burden, such as the size of a tumor or the number of cancer cells, serves as an important biomarker for evaluating a patient's response to oncology treatments. The synthesis of longitudinal tumor
burden data with survival outcomes is a key aspect of evaluating treatment effects, especially at the early stage of clinical studies. Unfortunately, due to the complexity of the models involved, methods developed to synthesize longitudinal and survival data often require either computationally intensive numerical techniques for parameter estimation or a substantial amount of data \cite{rizopoulos2012joint}. Specifically, when the aim is to
efficiently quantify the signal of treatment effectiveness in early-phase cancer studies, the scarcity of data poses a significant challenge in developing a framework to evaluate the treatment effect. Motivated by this issue, we propose a new Bayesian approach to construct a utility-based multi-component endpoint that unifies tumor burden outcomes with
survival outcomes for treatment effect evaluation.

The event of interest in our framework is death or disease progression. The time-to-event (survival) outcome is therefore defined as the progression-free survival (PFS) time. PFS and overall survival (OS) are two widely used endpoints for evaluating the clinical benefits of a drug. PFS, typically used as a surrogate measure of OS, represents the time from treatment initiation to disease progression. OS, on the other hand, serves as a direct measure of clinical benefit, indicating the duration of patient survival from the time of treatment initiation. Mathematically, OS can be decomposed into PFS and post-progression survival. The relationship between tumor burden and disease progression has been established in the RECIST guideline \citep{eisenhauer2009new}, which defines progression in the target lesion as a relative increase of $20\%$ in the sum of diameters of target lesions from the smallest sum observed during the study and an absolute increase of at least $5$ mm.
Although OS is considered the gold standard for demonstrating clinical benefit in cancer-related drugs during phase III confirmatory studies \citep{driscoll2009overall, zhuang2009overall, wilson2015outcomes}, using OS as the primary efficacy endpoint has certain limitations \citep{dejardin2010joint}: it requires large studies
with long periods of follow-up and is potentially confounded by other terminating events (e.g., non-cancer deaths). Moreover, if data on biomarkers such as tumor burden are available in the study, it is sensible to integrate
this information into treatment effect evaluation. Therefore, it is desirable to develop a single estimand that combines longitudinal biomarker data
and survival endpoint data to quantify treatment effectiveness in these studies.

The joint analysis of longitudinal and survival data is a well-explored area of research \citep{rizopoulos2012joint, crowther2013joint,lawrence2015joint}. The fundamental idea behind joint modeling is to incorporate longitudinal information into a time-to-event outcome model by factorizing the joint likelihood into two components: i) a longitudinal component to model the trajectory of longitudinal measurements and ii) a survival component to model the time-to-event occurrence. The standard practice is to either factorize the joint likelihood in two stages, first modeling the longitudinal outcome and then
conditioning the time-to-event data on these longitudinal outcomes, or to use shared random effects in conditionally independent models for the longitudinal and
time-to-event data. In our setting, we factorize the joint distribution into a time-to-event component to model the survival data and a longitudinal component, given the survival data, to model the distribution of the longitudinal tumor burden data conditional on a specific event occurring at a particular time. Such a factorization is less common in practice in the joint modeling literature but has been previously
explored \cite{hogan1997mixture}.

The clinical objective driving our work is to assess the treatment effect on both tumor burden and survival outcomes. To achieve this, we propose a multicomponent utility-based endpoint that integrates these outcomes, following
the composite variable strategy outlined in the FDA ICH E9 (R1) guidance \citep{ich2020e9}. On the one hand, observations after death are unavailable for any
subject. On the other hand, it may not always be feasible to determine whether a subject who died had, or would have had, a suspected disease. In such a case, defining the variable as a composite of death or disease progression may offer more insights
or facilitate the ascertainment of disease diagnosis before death. We construct a utility function that is a composite of tumor burden and survival outcomes, utilizing the area-under-the-curve (AUC) to reflect the planned follow-up duration. 

The article is structured as follows. In Sections \ref{sec2} and \ref{sec3}, we introduce our notation and define the utility-based composite strategy. Section \ref{sec4} describes our proposed joint modeling approach for longitudinal and survival data. Section
\ref{sec5} outlines the methodology for treatment effects estimation and inference. In Section \ref{sec6}, we conduct a simulation study to analyze the frequentist characteristics of the proposed estimand. We conclude our paper with a discussion in Section \ref{sec7}.

\section{Notation and setup}\label{sec2}

Consider a setting of a randomized cancer clinical trial with $N$ patients. For patient $i$, suppose that tumor burden assessments are scheduled at $n_i$ post-treatment visits. Each patient's information may be observed at a potentially different number of visits. Let $\textbf{X}_i$ be a collection of baseline pre-treatment covariates and $Y_{ij}$ be the tumor burden measurement observed at visit $t_{ij}, j \in \{0, \cdots, n_i\}$ (where $t_{i0} = 0$ corresponds to the baseline measurement time). For each patient, we have a true underlying time-to-event (time to death or progression), $T_i$, and a censoring time, $C_i$. The pair ($T_i, C_i$) defines the observed survival time, $S_i = min(T_i, C_i)$.  For patient $i$, $S_i$ is the minimum of their event time and censoring time. The event indicator $\delta_i$, defined as $\delta_i = I\big(T_i<C_i\big)$, distinguishes the event ($\delta_i = 1$) from censoring ($\delta_i = 0$). We let $L_i$ be the time from randomization to the analysis (e.g., time to the end of the study).  Similarly, let $A_i$ denote the treatment assignment indicator that distinguishes the treatment arm ($A_i =1$) from the control arm ($A_i =0$). 
At time $t$, we define $m_i(t)$ as the true tumor burden process. The observed tumor burden $Y_{ij}$ is related to the true underlying process through the equation:
\begin{equation}
    Y_{ij} = m_i(t_{ij}) + \epsilon_{ij},
\end{equation}
where $\epsilon_{ij}$ is a random error with mean zero. Details on the form of the true tumor burden process, $m_i(\cdot)$, are provided in Section \ref{sec42}. Using this notation, the observed data for patient $i$ is:
\begin{equation}
    \begin{aligned}
        \mathcal{O}_{i} &= \big\{\textbf{X}_{i}, A_{i}, \{Y_{i}(t_{ij}): j = 1, ..., n_{i}\}, L_i, S_{i}, \delta_{i} \big\}.
    \end{aligned}
\end{equation}

\section{Utility function}\label{sec3}
The motivation for using a utility function arises in settings where the outcome of interest for evaluating treatment regimes is either more complex than a binary or continuous random variable or, as in our case, involves the joint analysis of multiple random variables measured in distinct units. For patient $i$, we define the utility function as follows:
\begin{equation}
    u_{i}(t) = \begin{cases}
        m_{i}(t)  & t \leq T_{i}\\
        \Gamma & t> T_{i}. \\
    \end{cases}
\end{equation}
where $\Gamma > 0$ is a utility value corresponding to death or progression. In the utility function $u_{i}(t)$, the specification $m_{i}(t)$ \big(given $t \leq T_{i}$ and missing at random (MAR) assumption\big) can be data-driven. The MAR assumption \citep{rubin1976inference} asserts that the data missingness mechanism depends only on the observed values and not on the values of the data that are missing.  The specification of $\Gamma$, i.e. the utility post-event, is based on the clinical question of interest. 

For patient $i$, the utility-based endpoint is defined as:
\begin{equation}
    U_{i}= \int_{0}^{L_i}u_{i}(t) dt.
\end{equation}
The endpoint $U_{i}$ can be viewed as the "area under the curve” (AUC) for the utility function in the time interval $[0, L_i]$. Using $U_{i}$,  we define our parameter of interest, the average treatment effect (ATE), as:
\begin{equation}\label{eq5}
    ATE = E\big[U_{i} \big|A = 1\big] - E\big[U_{i} \big|A = 0\big] .
\end{equation}
Notice that the ATE in (\ref{eq5}) is not defined as the contrast between the conditional population averages of the longitudinal outcome or the survival outcome given treatment status. Instead, it is defined as the difference between the conditional expectations of the composite score $U_{i}$, which synthesizes the treatment effect on both (longitudinal and survival) outcomes.

The endpoint $U_i$ can be decomposed into the AUC corresponding to the longitudinal outcome and to the survival outcome. Intuitively, the average AUC corresponding to the survival outcome is the product of the post-event utility value ($\Gamma$) and one minus the restricted mean survival time $(RMST)$, with the cutoff being the average time on study. $RMST$ is defined as the area under the Kaplan–Meier survival curve up to a specific time point \citep{irwin1949standard,uno2014moving}. For a random variable $T$, the $RMST$ is the mean of the survival time ($min(T, t^{*})$) limited to some time horizon $t^{*}$ \citep{royston2013restricted}:
$$RMST = E\big[min(T, t^{*})\big].$$ It can be interpreted as a measure of average survival time or life expectancy from time 0 to a specific follow-up point, usually the event or censoring time. For example, when $T$ is years to death, we may think of $RMST$ as the ‘$t^{*}$-
year life expectancy.' Since $\Gamma$ is defined as the post-event utility value in our setting, it corresponds to the area above the Kaplan–Meier survival curve, which is given by $1 - RMST$. Therefore, the total treatment effect (i.e., the treatment effect on both tumor burden and survival outcomes) becomes:
\begin{align*}
   E\Big[\int_{0}^{min(L_i,T_i)}m_i(t) dt\Big]+ \Gamma \times \big(1 - RMST\big).
\end{align*}

Note that, in practice, the choice of the post-event utility value $\Gamma$ is subjective and typically guided by expert advice or clinical relevance. The available data consists of tumor burden measurements collected at various time points. When an event, such as death or disease progression, occurs, the post-event utility value $\Gamma$ reflects the additional penalty assigned to that event. Specifically, $\Gamma$ translates to a $\Gamma \times 100\%$ increase in the contribution of the event to the overall utility-based endpoint, compared to the level of contribution of the tumor burden before the event’s occurrence.
For example, if $\Gamma=2$, this means that an event is assigned a $200\%$ penalty in computing $U_i$, relative to the tumor burden measurements alone. In essence, $\Gamma$ acts as a weight that quantifies how much more impactful the occurrence of death or disease progression is relative to the longitudinal tumor burden trajectory alone. 

It is important to recognize that as $\Gamma$ becomes larger, it increasingly dominates the calculation of the composite endpoint $U_i$. Consequently, the contribution of the longitudinal tumor burden measurements to $U_i$ becomes less significant in the presence of a large post-event penalty. This highlights the need for careful selection of $\Gamma$ based on clinical context and the importance of balancing the survival and longitudinal aspects of the data within the composite endpoint. In our simulation study (details in Section \ref{sec6}), we set $\Gamma=0.5$, indicating a $50\%$ penalty for death or disease progression when computing $U_i$, which strikes a balance between adequately accounting for the contribution of survival outcomes and preserving the contribution of longitudinal tumor burden measurements.

\section{Model specification}\label{sec4}

In this article, we factorize the joint model into (i) a survival sub-model and (ii) a longitudinal sub-model conditional on the survival data, and model them separately. In doing so, we assume that, given the observed covariate history, the censoring mechanism and the visiting process (the mechanism that determines the time points at which longitudinal tumor burden measurements are collected) are independent of the true event times and future longitudinal measurements. These assumptions imply that decisions about whether a subject withdraws from the study or appears for a longitudinal tumor burden measurement depend only on the observed covariate history (previous longitudinal and baseline measurements).

\subsection{Survival model} \label{sec41}

The proposed approach can accommodate any survival regression model. However, we prefer a model with convenient properties that facilitate prediction and data augmentation. Therefore, assuming censoring is non-informative, we consider a parametric exponential model for our time-to-event outcome, $T_i$. Suppose $\lambda_{i}$ denotes the instantaneous hazard of experiencing an event (death or disease progression) for patient $i$. We define $\lambda_{i}$ as a function of the treatment indicator $A_i$ and pre-treatment covariates $\textbf{X}_i$ as:
\begin{equation}
    \lambda_{i} = \exp(\gamma_{A} A_i + \boldsymbol{\gamma}^{T}_{\textbf{X}} \textbf{X}_i).
\end{equation}
Finally, we assume that the time-to-event outcome for patient $i$ follow an exponential distribution with hazard $\lambda_{i}$, given by:
\begin{equation}
    T_i \sim Exponential(\lambda_i).
\end{equation}

\subsection{Longitudinal model given survival outcome} \label{sec42}
We define  $\psi_{ij} = \frac{t_{ij}}{T_{i}} $ as the scaled visit time for modeling the true tumor burden process, denoted by $m_{i}(\psi_{ij})$. By definition, $\psi_{ij}$ lies between $0$ and $1$, i.e., $\psi_{ij} \in [0,1]$ for each subject. This scaled visit time, which is a deterministic function of the true event time $T_i$, enables modeling the tumor burden trajectory for each subject on a common time scale between $0$ and $1$, rather than on a subject-specific time scale between $0$ and $T_i$. Let $\textbf{b}_{i} \equiv \begin{pmatrix}
b_{1,i} \\
b_{2,i}
\end{pmatrix}$ be a random-effects vector associated with subject $i$. The observed longitudinal outcome can be expressed as:
 \begin{equation}
        Y_{ij}\big|\textbf{b}_i,\textbf{X}_{i}, A_{i}, T_{i} = m_{i}(\psi_{ij}) + \epsilon_{ij},
    \end{equation}
where $E[\epsilon_{ij}] = 0$, $Var[\epsilon_{ij}] = \sigma^{2}$, and $Cov[\epsilon_{ij},\epsilon_{ij'}] = 0$ if $j \neq j'$. Note that $T_{i}$ impacts $Y_{ij}$ only through $\psi_{ij}$. We model $m_{i}(\psi_{ij})$ with a linear mixed model with a quadratic effect component for scaled time:
\begin{equation} \label{meanFuncTBmodel}
    \begin{aligned}
        m_{i}(\psi_{ij})  &= \boldsymbol{\beta}^{T}_{\textbf{X}}\textbf{X}_{i}  + \beta_{A} A_{i} +  b_{1,i}\psi_{ij} +  b_{2,i}\psi_{ij}^{2},
    \end{aligned}
\end{equation}
where $ b_{1,i} \text{ and
} b_{2,i}$ are assumed to follow bivariate normal distribution, i.e.,
\[\begin{pmatrix}
b_{1,i} \\
b_{2,i}
\end{pmatrix}\sim MVN \left(\begin{pmatrix}
b_{1} \\
b_{2}
\end{pmatrix},\Sigma = \begin{pmatrix}
\sigma_{11} & \sigma_{21} \\
\sigma_{12} & \sigma_{22}
\end{pmatrix}\right),
\]
with hyperparameters $b_{1},b_{2} \text{ and } \Sigma$. We highlight that the treatment effect estimation procedure described in this article is not tied to the specific form of the mean function (\ref{meanFuncTBmodel}) and the utility-based design remains valid even if a different model specification is used. We may also define our longitudinal outcome at visit $j$ as the change in tumor burden from baseline. For example, at each visit $j$ we can define the  percentage change in observed tumor burden \big($Y_{ij}^{'}$\big) from the baseline measurement, \big($Y_{i0}$\big) , as:
\begin{equation}
    \begin{aligned}
        Y_{ij}^{'}\big|\textbf{b}_i,\textbf{X}_{i}, A_{i}, T_{i} &=  \frac{m_{i}(\psi_{ij}) + \epsilon_{ij} - Y_{i0}}{Y_{i0}}.
    \end{aligned}
\end{equation}

\subsection{Likelihood and posterior computation}\label{sec43}
Let $\xi = \big(\gamma_{A}, \boldsymbol{\gamma}_{\textbf{X}}, \boldsymbol{\beta}_{\textbf{X}}, \beta_{A}, \textbf{b}_{i}, \textbf{b}, \sigma^{2}, \Sigma\big)$ denote all the parameters from the two sub-models. Define $\xi_{S} = \big(\gamma_{A}, \boldsymbol{\gamma}_{\textbf{X}}\big)$ and $\xi_{Y} = \big(\boldsymbol{\beta}_{\textbf{X}}, \beta_{A}, \textbf{b}_{i}, \textbf{b}, \sigma^{2}, \Sigma\big)$ as the parameters from the survival and longitudinal sub-models, respectively. Thus, we have $\xi = \big(\xi_{S}, \xi_{Y}\big)$. The joint likelihood of the parameters is then given by:
\begin{equation}
    \begin{aligned}
        L\big(\xi; \textbf{Y},\textbf{X},\textbf{A},\textbf{S}, \boldsymbol{\delta}\big) 
        & \propto
        \prod_{i: \delta_i =1} p\big(\textbf{Y}_i \big| S_i, \textbf{X}_i, A_i; \xi_{Y}  \big) p\big(S_i \big|  \textbf{X}_i, A_i; \xi_{S}\big) 
         \prod_{i: \delta_i =0} \int_{S_i}^{\infty} p\big(\textbf{Y}_i \big| t, \textbf{X}_i, A_i; \xi_{Y} \big) p\big(t \big|  \textbf{X}_i, A_i; \xi_{S} \big) dt
         \\&
         = \prod_{i: \delta_i =1} \bigg[\prod_{j =1}^{n_i} p\big(Y_{ij} \big| S_i, \textbf{X}_i, A_i; \xi_{Y}  \big)\bigg] p\big(S_i \big|  \textbf{X}_i, A_i; \xi_{S}\big) 
         \prod_{i: \delta_i =0} \int_{S_i}^{\infty} \bigg[\prod_{j =1}^{n_i} p\big(Y_{ij} \big| t, \textbf{X}_i, A_i; \xi_{Y} \big)\bigg]
         p\big(t \big|  \textbf{X}_i, A_i; \xi_{S} \big) dt
    \end{aligned}
\end{equation}
where $\textbf{S} = \big[S_1, \ldots, S_N\big]^\top$, $\textbf{A} = \big[A_1, \ldots, A_N\big]^\top$, $\textbf{Y} = \big[\textbf{Y}_1, \ldots, \textbf{Y}_N\big]^\top$ with $\textbf{Y}_i = \big[Y_{i,1}, \ldots, Y_{i, n_i}\big]$, and $\textbf{X} = \big[\textbf{X}_1, \ldots, \textbf{X}_N\big]^\top$ with $\textbf{X}_i = \big[X_{i,1}, \ldots, X_{i, k}\big]$ for $i = 1, \ldots, N$ and $k$ baseline covariates.  The joint posterior distribution of the unknown parameter $\xi$, denoted $p\big(\xi|\textbf{Y},\textbf{X},\textbf{A},\textbf{S}\big)$, has the following form:
\begin{equation}
    \begin{aligned}
      p\big(\xi \big|\textbf{Y},\textbf{X},\textbf{A},\textbf{S}\big) &\propto L\big(\xi;\textbf{Y},\textbf{X},\textbf{A},\textbf{S}\big) \times p\big(\xi\big)
    \end{aligned}
\end{equation}
where $p\big(\xi\big)$ represents the joint prior distribution of the parameters. We assume (without loss of generality) that the parameters are a-priori independent.

\subsection{Data augmentation}\label{sec44}
Recall that the utility end-point is given by:
\begin{equation}
    \begin{aligned}
        U_{i}
              &= \int_{0}^{min\{L_{i},T_{i}\}} m_{i}(t) dt  +
              \int_{T_{i}}^{L_{i}} \Gamma \times I\big(T_{i} < L_{i}\big) dt \\
              &= \int_{0}^{min\{L_{i},T_{i}\}} m_{i}(t) dt  +
              \Gamma \times  \big(L_{i} - T_{i}\big) \times I\big(T_{i} < L_{i}\big).
    \end{aligned}
\end{equation}
For individuals with $T_i < \min(C_i, L_i)$ (i.e., individuals who experience death or disease progression while they are in the study), the true event time ($T_{i}$) is observed. For notational clarity, we denote this event time as $T_{i, \text{obs}}$. For individuals who did not experience death or disease progression until the end of the study, the true event time ($T_{i}$) is unknown. For these individuals (whose true event time is longer than the end-of-study time), we treat $T_{i}$ as missing data and impute it using data augmentation. Specifically, defining $T_{i, \text{miss}}$ as the event time of individuals who were censored at $C_{i}$, right-censoring is handled by data augmentation as follows:
\begin{equation}
    T_{i}
        \begin{cases}
             = T_{i, obs} \quad & \text{if individual $i$ experiences event} \\
             \sim p(T_{i, miss}|Y_{i},S_{i},T_i>C_i, A_{i}, \textbf{X}_{i}) = \frac{p(T_{i},Y_{i}|S_{i},T_i>C_i, A_{i}, \textbf{X}_{i})}{p(Y_{i}|S_{i},T_i>C_i, A_{i}, \textbf{X}_{i})}  \quad & \text{if individual $i$ is censored.}

        \end{cases}
\end{equation}

\section{Inference}\label{sec5}
We sample from the posterior distributions of the parameters using Stan \citep{carpenter2017stan}. Stan uses Hamiltonian Monte Carlo (HMC) \citep{neal2012mcmc,betancourt2017conceptual} to update the parameters simultaneously, collapsing Steps 2 (a) and (b) in Algorithm \ref{alg:1} into a single step,
\begin{align*}
    p\big(\xi^{(q)}, T_{i, miss}^{(q)} \big|\textbf{Y},\textbf{S}, \boldsymbol{\delta}, \textbf{A},\textbf{X}\big) &\propto p\big(\textbf{Y},\textbf{S}, \boldsymbol{\delta}, \textbf{A},\textbf{X} \big|\xi, T_{i, miss}\big) \times p\big(\xi\big).
\end{align*}

\subsection{Treatment effect estimation}\label{sec51}

Recall that our goal is to estimate the ATE, where $ATE = E\big[U_{i}\big|A = 1\big] - E\big[U_{i}\big|A = 0\big]$. Denoting $\theta_{a} = E\big[U_{i}\big|A = a\big]$ and $\theta = ATE$, we rewrite the treatment effect as $\theta = \theta_{1} - \theta_{0}$. Let $\hat{\theta}_{a}$ denote our estimator of $\theta_{a}$, defined as:
\begin{equation}
    \begin{aligned}
        \hat{\theta}_{a} &= \frac{1}{Q} \sum_{q=1}^{Q} \hat{\theta}^{(q)}_{a} 
        = \frac{1}{Q} \sum_{q=1}^{Q} \bigg\{ \frac{1}{n_{a}} \sum_{i=1}^{n_{a}} I_{\{A_{i} = a \}} U_{i}^{(q)}\bigg\},
    \end{aligned}
\end{equation}
where $U_{i}^{(q)}$ denotes the posterior sample of $U_{i}$ at iteration $q \in \{1, \ldots, Q\}$, and  $n_{1}$ and $n_{0}$ denote the sample sizes of the treatment arm and the control arm, respectively. 

We compute $U_{i}^{(q)}$ using the posterior sample
$\xi^{(q)} = \big(\gamma_{A}^{(q)}, \boldsymbol{\gamma}_{\textbf{X}}^{(q)}, \boldsymbol{\beta}_{\textbf{X}}^{(q)}, \beta_{A}^{(q)}, \textbf{b}_{i}^{(q)},\textbf{b}^{(q)}, \sigma^{2,(q)}, \Sigma^{(q)}\big)$ (sampling details are available in Algorithm \ref{alg:1}). In particular, we calculate the area under the curve for each individual $i$ from $0$ to $\min(L,T_i^{(q)})$; for patient $i$ we calculate the area under the tumor burden curve, denoted as $TBAUC_{i}^{(q)}$:

\begin{equation}
    \begin{aligned}
      TBAUC_{i}^{(q)} &= \int_{0}^{\min(L,T_i^{(q)})} m_i\Big(t \big|\beta_{X}^{(q)},\beta_{A}^{(q)}, b_{1i}^{(q)}, b_{2i}^{(q)} \Big) dt 
      = \int_{0}^{\min(L,T_i^{(q)})} \Big(\beta_{X}^{(q)}X_{i} +  \beta_A^{(q)} A_i + b_{1i}^{(q)} \psi_{ij} + b_{2i}^{(q)} \psi_{ij}^2 \Big) dt \\
      &= \int_{0}^{\min(L,T_i^{(q)})} \Big(\beta_{X}^{(q)}X_{i} +  \beta_A^{(q)} A_i + b_{1i}^{(q)} \frac{t}{T_i^{(q)}} + b_{2i}^{(q)} \frac{t^2}{\big(T_i^{(q)}\big)^2}\Big) dt \\
      &=  [t]_{0}^{\min(L,T_i^{(q)})} 
      \Big\{\beta_{X}^{(q)} X_{i}  +  \beta_A^{(q)} A_i  \Big\}  + b_{1i}^{(q)} \frac{ [t^2/2]_{0}^{\min(L,T_i^{(q)})}}{T_i^{(q)}} + b_{2i}^{(q)} \frac{[t^3/3]_{0}^{\min(L,T_i^{(q)})}}{\big(T_i^{(q)}\big)^2}  \\
      &=  \min(L,T_i^{(q)}) \Big\{\beta_{X}^{(q)}X_{i}  +  \beta_A^{(q)} A_i  \Big\}  + b_{1i}^{(q)} \frac{\min(L,T_i^{(q)})^2}{2T_i^{(q)}} + b_{2i}^{(q)} \frac{\min(L,T_i^{(q)})^3}{3\big(T_i^{(q)}\big)^2}.
    \end{aligned}
\end{equation}
Note that if $\min(L_i,T_i^{(q)}) = L_i$, then we must integrate $TBAUC_{i}^{(q)}$ up to $L_i$ and the area under the survival curve, denoted $SAUC_{i}^{(q)}$, is $0$. However, if $\min(L, T_i^{(q)}) = T_i^{(q)}$, then $SAUC_{i}^{(q)} = \big(L_i -\min(L_i,T_i^{(q)})\big) \times \Gamma$ where $\Gamma$ is the utility value corresponding to death or disease progression. These two quantities together decompose the total area under the curve,
\begin{equation}\label{eq:3}
    U_{i}^{(q)} = TBAUC_{i}^{(q)} + SAUC_{i}^{(q)}.
\end{equation}
The corresponding estimator of $\theta$, say $\hat{\theta}$, given by:
\begin{equation}\label{eq::thetaHat}
    \begin{aligned}
        \hat{\theta} = \hat{\theta}_{1} - \hat{\theta}_{0} = \frac{1}{Q} \sum_{q = 1}^{Q} \bigg\{\frac{1}{n_{1}} \sum_{i=1}^{n_{1}} I_{\{A_{i} = 1 \}} U_{i}^{(q)} -  \frac{1}{n_{0}} \sum_{i=1}^{n_{0}} I_{\{A_{i} = 0 \}} U_{i}^{(q)}\bigg\}.
    \end{aligned}
\end{equation}
The steps of the algorithm to compute the posterior samples of $\hat{\theta}$ are given in Algorithm \ref{alg:1}. 

\begin{algorithm}
\caption{Algorithm  to compute posterior samples of  $\hat{\theta}_{a}$}\label{alg:1}
\begin{enumerate}
    \item  For $q \in \{1,...,Q \}$, define $\xi^{(q)} = (\xi^{(q)}_{S},\xi^{(q)}_{Y})$ as the $q^{th}$ posterior draw of the parameters. Here, $ \xi^{(q)}_{S} = \big(\gamma_{A}^{(q)}, \boldsymbol{\gamma}_{\textbf{X}}^{(q)}\big) $ and $\xi^{(q)}_{Y} = \big(\boldsymbol{\beta}_{\textbf{X}}^{(q)}, \beta_{A}^{(q)}, \textbf{b}_{i}^{(q)},\textbf{b}^{(q)}, \sigma^{2,(q)}, \Sigma^{(q)}\big)$. Then, set $\xi^{(0)} \gets $ initial value of the parameters.
    \item For $q \in \{1,...,Q \}$ sequentially do the following:
            \begin{enumerate}
                    \item Sample new value of parameters from the posterior distribution:
                    \begin{align}
                        p\big(\xi^{(q)}\big|\textbf{Y},\textbf{S}, \boldsymbol{\delta}, \textbf{A},\textbf{X}\big) &\propto p\big(\textbf{Y},\textbf{S}, \boldsymbol{\delta} \big|\textbf{A},\textbf{X},\xi\big) \times p\big(\xi\big)\\
                        &= \prod_{i=1}^{N} \prod_{j=1}^{n_{i}} p\big( S_{i}, \delta_{i} \big| A_i, X_i, \xi_{S}\big)\times p\big( Y_{ij} \big| S_{i}, \delta_{i}, A_i, X_i, \xi_{Y}\big) \times p\big(\xi\big).
                    \end{align}
                    \item Sample true event times for censored subjects from the posterior distribution:
                    \begin{align}
                        p\big(T_{i, miss}^{(q)}\big|\textbf{Y},\textbf{S}, \boldsymbol{\delta}, \textbf{A},\textbf{X},\xi^{(q)}\big) &\propto  \prod_{i=1}^{N} \prod_{j=1}^{n_{i}} p\big( S_{i}, \delta_{i}\big|A_i, X_i, \xi_{S}^{(q)}\big)\times p\big( Y_{ij} |S_{i}, A_i, X_i, \delta_{i}, \xi_{Y}^{(q)}\big) \times p\big(T_{i,miss}\big).
                    \end{align}
            \end{enumerate}

    \item For each individual $i$, compute $Q$ posterior samples of $U_{i}^{(q)}$, $q \in \{1, ..., Q \}$, using Equation (\ref{eq:3}) and the posterior samples from Step 2.
    \item At each iteration $q \in \{1,...,Q \}$, compute $\hat{\theta}^{(q)}_{a} =  \frac{1}{n_a}\sum_{i=1}^{n_{a}} I_{\{A_{i} = a \}} U^{(q)}_{i}$ for $a \in \{0,1\}$.
    \item Compute  $\hat{\theta}_{a} = \frac{1}{Q} \sum_{q=1}^{Q} \hat{\theta}^{(q)}_{a}$.
\end{enumerate}
\end{algorithm}

\subsection{Hypothesis testing and uncertainty quantification}\label{sec52}

Suppose $\theta_1$ and $\theta_0$ denote the population average total area under the curve (AUC) for subjects in the treatment and the control arms, respectively. For $A = a$, let $\theta^{TB}_a$ and $\theta^{S}_a$ represent the population average AUC within each treatment stratum for the longitudinal and survival components, respectively. The arm-specific differences between these population AUC averages result in the true treatment effects, denoted $\theta$, $\theta^{TB}$ and $\theta^{S}$, on the total AUC, TBAUC, and SAUC, respectively. 

The estimators for the population variances are computed as posterior variances of weighted sample average treatment effects across $Q$ iterations. Each iteration's weighted sample average treatment effect is computed using the Bayesian Bootstrap (BB) \citep{rubin1981bayesian} to avoid underestimating the variance of the AUC, which can occur when using the empirical mean. This variance underestimation can be particularly severe for SAUC, as the SAUC value is zero across iterations for subjects who die before the end of the study. Unlike the empirical mean, which assigns a uniform weight $1/n_a$ to each $I_{\{A_{i} = a \}} \times AUC^{(q)}_i$, BB assumes the weights $w_{a,i,BB}$ are unknown parameters. 

For each $a \in \{0,1\}$, the weights $w_{a,i,BB}$ live in the simplex $\mathcal{W}_{a, BB} \equiv \big\{\mathbb{R}^{n_a}: w_{a,i,BB} > 0 \quad  \forall i \in \{i: A_i = a\}, \sum_{j=1}^{n_a} w_{a,j,BB} =1 \big\}$.
The BB assigns an improper Dirichlet prior, $w_{a,i,BB} \sim Dirichlet \big( 0_{n_a}\big)$ ,over $\mathcal{W}_{a, BB}$ where $0_{n_a}$ is the $n_a$ dimensional zero vector. This prior is conjugate, yielding a posterior distribution $w_{a,i,BB}\big|\big\{ AUC^{(q)}_i: A_{i} = a \big\} \sim Dirichlet \big( 1_{n_a}\big)$, where $1_{n_a}$ is the $n_a$ dimensional vector of ones. 

Bootstrap inference is often used to quantify treatment effect estimates in causal inference problems \citep{imbens2004nonparametric, otsu2017bootstrap, jiang2024bootstrap}. However, these studies employ bootstrap procedures to compute both point and interval estimates of the average treatment effects. In our work, the point and interval estimates of the average treatment effects are derived from the posterior mean and variance of ATEs across $Q$ iterations. We use BB as an intermediate step to estimate the posterior variance for the Wald-type significance tests.

The estimators for the true treatment effect on TBAUC and its variance are defined as:
\begin{align*}
    \hat{\theta}^{TB}  &= \frac{1}{Q} \sum_{q = 1}^{Q} \bigg\{\frac{1}{n_{1}} \sum_{i=1}^{n_{1}} I_{\{A_{i} = 1 \}} TBAUC_{i}^{(q)} -  \frac{1}{n_{0}} \sum_{i=1}^{n_{0}} I_{\{A_{i} = 0 \}} TBAUC_{i}^{(q)}\bigg\},
    \\
    SE\big(\hat{\theta}^{TB}\big)^2 &= Var_q\bigg( \Big\{ \sum_{i=1}^{n_{1}} w^{TB}_{1,i,BB} \big(I_{\{A_{i} = 1 \}} \times  TBAUC_{i}^{(q)}\big) -  \sum_{i=1}^{n_{0}} w^{TB}_{0,i,BB} \big( I_{\{A_{i} = 0 \}} \times  TBAUC_{i}^{(q)}\big)\Big\}\bigg) ,
\end{align*}
and the estimators for the true treatment effect on SAUC and its variance are similarly defined as:
\begin{align*}
    \hat{\theta}^{S}  &= \frac{1}{Q} \sum_{q = 1}^{Q} \bigg\{\frac{1}{n_{1}} \sum_{i=1}^{n_{1}} I_{\{A_{i} = 1 \}} SAUC_{i}^{(q)} -  \frac{1}{n_{0}} \sum_{i=1}^{n_{0}} I_{\{A_{i} = 0 \}} SAUC_{i}^{(q)}\bigg\}, \\
    SE\big(\hat{\theta}^{S}\big)^2  &= Var_q\bigg( \Big\{ \sum_{i=1}^{n_{1}} w^{S}_{1,i,BB}\big(I_{\{A_{i} = 1 \}} \times SAUC_{i}^{(q)}\big) -  \sum_{i=1}^{n_{0}} w^{S}_{0,i,BB}\big(I_{\{A_{i} = 0 \}} \times  SAUC_{i}^{(q)}\big)\Big\}\bigg).
\end{align*}
Finally, the estimator for the true treatment effect on total AUC is given by Equation \ref{eq::thetaHat}, and the estimator for the population variance is defined as:
\begin{align*}
    SE\big(\hat{\theta}\big)^2 &= Var_q\bigg( \Big\{ \sum_{i=1}^{n_{1}} w^{Tot}_{1,i,BB} \big( I_{\{A_{i} = 1 \}} \times  U_{i}^{(q)} \big)-  \sum_{i=1}^{n_{0}} w^{Tot}_{0,i,BB} \big( I_{\{A_{i} = 0 \}} \times  U_{i}^{(q)}\big) \Big\}\bigg).
\end{align*}
Table \ref{tab1} outlines our hypothesis testing setup, focusing on population-level treatment effects on TBAUC, SAUC, and total AUC. Using the estimators for these parameters, as defined in this section, we conduct one-sided significance tests for the null hypothesis of no treatment effects on these AUCs. 
\begin{center}
\begin{table*}[!h]%
\caption{Significance testing procedures.\label{tab1}}
\begin{tabular*}{\textwidth}{@{\extracolsep\fill}lllll@{}}
\toprule
\textbf{Area under the curve} & \textbf{Parameter}  & \textbf{Estimator}  & {\textbf{Hypotheses}}  & \textbf{Test statistic}   \\
\midrule
Longitudinal AUC only &
$\theta^{TB} = \theta^{TB}_1 - \theta^{TB}_0$  &
 $\hat{\theta}^{TB}$ & 
 $H_{0}^{TB}: \theta^{TB} = 0$   & 
 $W^{TB} = \frac{\hat{\theta}^{TB}}{SE(\hat{\theta}^{TB})}$   \\ \cmidrule{2-3}
                        & 
                        $Var\big( \theta^{TB} \big)$  &
                        $SE\big(\hat{\theta}^{TB}\big)^2$ &
                        $H_{A}^{TB}: \theta^{TB} > 0$   & 
                           \\
 \midrule
Survival AUC only & 
$\theta^{S} = \theta^{S}_1 - \theta^{S}_0$ &
 $\hat{\theta^{S}}$ & 
$H_{0}^{S}: \theta^{S} = 0$  &
$W^{S} = \frac{\hat{\theta^{S}}}{SE(\hat{\theta^{S}})}$   \\ \cmidrule{2-3}
                  & 
                  $Var\big(\theta^{S}  \big)$  &
                  $SE\big(\hat{\theta^{S}}\big)^2$  & 
                  $H_{A}^{S}: \theta^{S} > 0$ &
                    \\
 \midrule
Total AUC & 
$\theta = \theta_1 - \theta_0$ &
 $\hat{\theta}$ & 
$H_{0}^{Tot}:\theta = 0$  & 
$W^{Tot} = \frac{\hat{\theta}}{SE(\hat{\theta})}$   \\ \cmidrule{2-3}
            & 
            $Var\big( \theta\big)$  &
            $SE\big(\hat{\theta}\big)^2$  &
            $H_{A}^{Tot}:\theta > 0$   & 
              \\
\bottomrule
\end{tabular*}
\end{table*}
\end{center}

\section{Simulation study}\label{sec6}
  
Recall that we assume an exponential distribution for the time-to-event outcome. This approach implies a constant hazard over time, providing a mathematically tractable framework for power and sample size calculations in time-to-event analyses \citep{collett2023modelling}. Under this assumption, we determined that approximately $117$ events are needed to achieve $90\%$ power to detect a moderate treatment effect---assuming a hazard ratio of approximately $0.74$, where “moderate” reflects a clinically meaningful $26\%$ relative risk reduction. Consequently, we design the study to conclude after observing a total of $120$ events. Accruing $120$ PFS events exceeds this threshold, supporting the adequacy of our study’s statistical power. This calculation aligns closely with sample-size calculation methods described for the proportional hazards regression model \citep{schoenfeld1983sample}.

Furthermore, accounting for anticipated follow-up time and the natural course of the disease, we select a total sample size of $500$ patients to ensure that at least $120$ PFS events will be observed during the study. Subjects are enrolled at a rate of $50$ individuals per month. Each individual has a maximum of $8$ visits, scheduled $9$ weeks apart. This design ensures that the study is adequately powered to detect a moderate treatment benefit and to provide reliable results. The R code used to implement our simulation studies is available at \url{https://github.com/SBstats/TumorBurdenSurvJointModel}.

\subsection{Data Generation}\label{sec61}
\subsubsection{Baseline covariates and random effects}\label{sec611}
Each subject $i$ is assigned a  continuous covariate, $X_{i}$, which is a realization from Normal distribution $\big(X_{i} \sim N(6,0.25)\big)$. We randomly assign each
subject to control ($A_i = 0$) or treatment ($A_i = 1$) with randomization ratio $1:1$. The random effects are generated as random samples from the Normal distributions $b_{1,i}  \sim N(-10, 1)$ and $b_{2,i}  \sim N(11, 1)$.

\subsubsection{Time to event and censoring}\label{sec612}
We define the hazard parameter, $ \lambda_{i} = exp\big(\gamma_{X} X_{i} + \gamma_A A_i\big)$, with $\gamma_{X} = -1.2$ and $\gamma_A$ set according to the simulation scenarios in Section \ref{sec62}. We generate time-to-event as a realization from: $$T_{i} \sim Exponential\big(\lambda_{i}\big).$$  Note that the median of the exponential distribution with rate parameter $\lambda_i$ is $\frac{ln(2)}{\lambda_i}$. This means the median true event time in the treatment and control groups are $\frac{ln(2)}{exp\big(-1.2 X_{i} - \gamma_A\big)}$ and $\frac{ln(2)}{exp\big(-1.2 X_{i}\big)}$, respectively.  Finally, we generate censoring
times, $C_i$, as random samples from $Exponential(0.00128)$ which sets the median censoring time to approximately $540$ days (18 months), mimicking a typical phase II oncology study.

\subsubsection{Tumor Burden}\label{sec613}
 We generate the baseline TB measurement for each subject, as $Y_{i0}\sim N(15,0.5)$. Define $\psi_{ij} = \frac{t_{ij}}{S_{i}} $, $\psi_{ij} \in [0,1]$. Conditional on $X_{i}$ , $A_i$,  $S_i$, $b_{1,i}$, and $b_{2,i}$, the tumor burden process is simulated using the following mixed effects model:

\begin{equation}
    \begin{aligned}
        Y_{ij}|\textbf{b}_{i}, X_{i}, A_{i}, S_{i} &= m_{i}(\psi_{ij}) + \epsilon_{ij} \\
        &= \beta_{X} X_{i} +  \beta_{A} A_{i} +  b_{1,i}\psi_{ij} +  b_{2,i}\psi_{ij}^{2} +  \epsilon_{ij},
    \end{aligned}
\end{equation}
where $\beta_{X} = 1.5$ and $\beta_{A}$ is set according to the simulation scenarios in Section \ref{sec62}. In the specification above, we have $E[\epsilon_{ij}] = 0$, $Var[\epsilon_{ij}] = 1.5$, and $Cov[\epsilon_{ij},\epsilon_{ij'}] = 0$ if $j \neq j'$.

\subsection{Simulation scenarios}\label{sec62}
We assume that taking the treatment is advantageous. This is indicated by negative values of the treatment effect coefficients---a negative $\beta_A$ suggests that taking treatment reduces tumor burden and a negative $\gamma_A$ suggests that the treatment group subjects live longer compared to their control group counterparts. We consider four scenarios based on the presence (or the absence) of a treatment effect in longitudinal and survival models (Table \ref{tab2}).

\begin{center}
\begin{table*}[!h]%
\caption{Treatment effect scenarios for the simulations \label{tab2}}
\begin{tabular*}{\textwidth}{@{\extracolsep\fill}lllll@{}}
\toprule
\textbf{Scenario} & \textbf{Treatment effect on tumor burden}  & \textbf{Treatment effect on survival}    \\
\midrule
1      &    Yes ($\beta_A = -2.25$)  &   Yes ($\gamma_A = -0.75$)\\
         2      &   Yes ($\beta_A = -2.25$)    &  No ($\gamma_A = 0$) \\
         3      &   No ($\beta_A = 0$)   &  Yes ($\gamma_A = -0.75$)\\
         4      &  No ($\beta_A = 0$)   &  No ($\gamma_A = 0$)\\
\bottomrule
\end{tabular*}
\end{table*}
\end{center}

\begin{figure}
        \centering
        \includegraphics[scale=0.5]{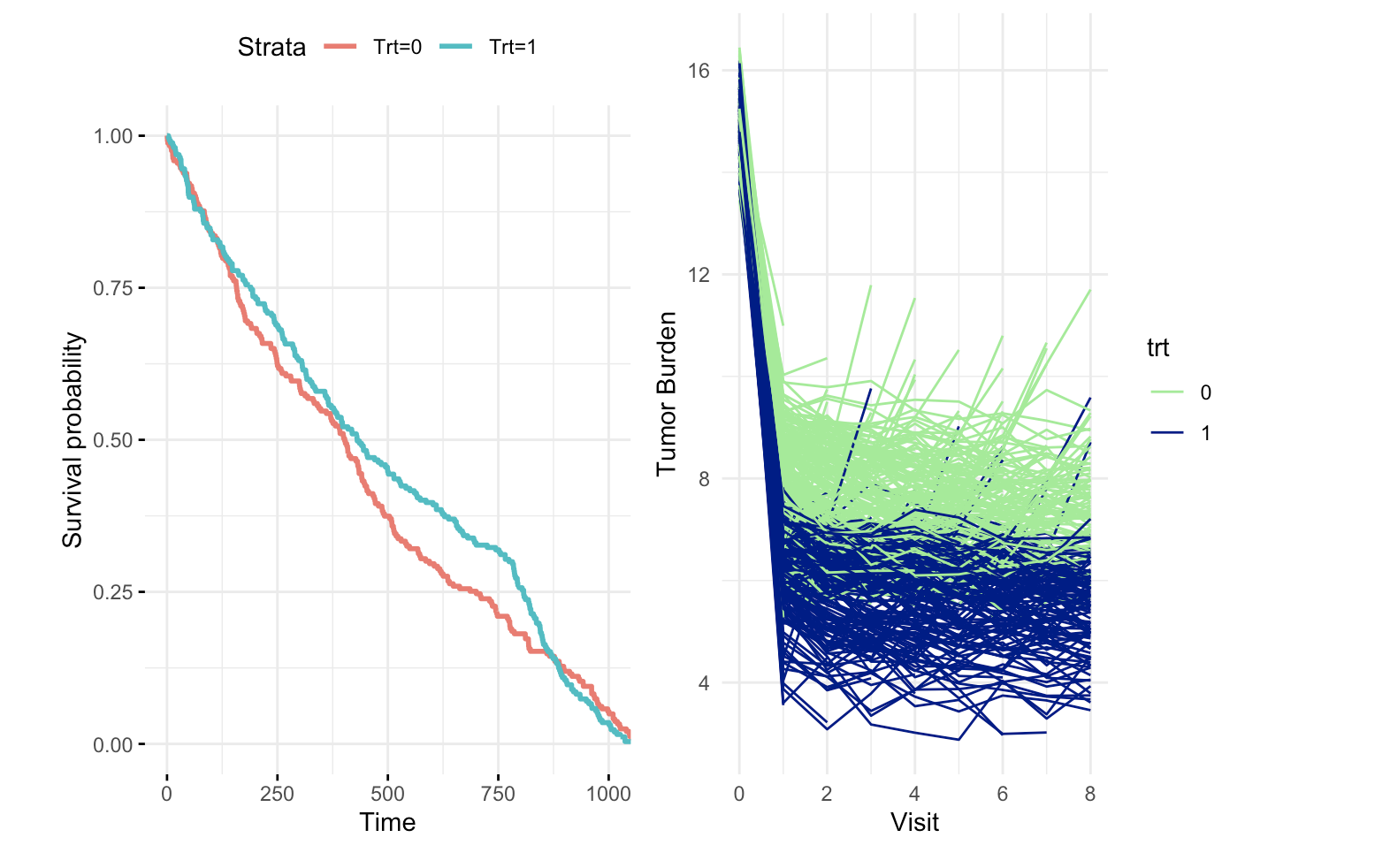}
        \caption{Survival outcome Kaplan-Meier (KM) curve and tumor burden outcome spaghetti plot for simulation scenario 1 from one data replication.}
        \label{fig:KMandTBcurveSim1}
\end{figure}

Figure \ref{fig:KMandTBcurveSim1} illustrates survival and longitudinal outcomes for simulation scenario 1 from a single data replication. Similar illustrations for scenarios 2-4 are available in Appendix \ref{AppendixB}.
For each simulation scenario, we have three hypothesis testing procedures: test for longitudinal area under the curve only, test for survival area under the curve only, and test for total area under the curve. Recall that to compute the total area under the curve, we set the post-event utility value, $\Gamma = 0.5$. Ideally, all three hypothesis testing procedures from Section \ref{sec5} should control for Type I error.

\subsection{Results}\label{sec63}
Table \ref{tab3} reports the rejection rates for one-sided (left-tailed) significance tests under the null hypotheses of no treatment effects on tumor burden AUC, survival AUC, and total AUC, respectively. In Scenario 4, where there are no treatment effects in both longitudinal and survival models, we expect the rejection rates for all three tests to be around $0.025$. As shown in Table \ref{tab3}, the rejection rate for testing the treatment effect on survival AUC in Scenario 4 is close to this value. However, for testing the effects on tumor burden and total AUC, the rejection rates are more conservative.

In Scenario 3, where no treatment effect exists in the longitudinal model, we would expect the rejection rate for the one-sided test under the null hypothesis of no treatment effect on tumor burden AUC to be around $0.025$. However, Table \ref{tab3} shows a higher-than-expected rejection rate of $0.062$ for the tumor burden AUC. This is due to the joint distribution of longitudinal and survival data being decomposed into a survival model and a longitudinal model conditional on survival data. As a result, if some treatment effect exists in the survival model (i.e., $\gamma_A < 0$), this effect can propagate into the longitudinal model, even when no treatment effect is present in the latter (i.e., when $\beta_A = 0$).
\begin{center} 
\begin{table*}[!h]%
\caption{Rejection rates for Wald-type tests across 1000 simulation replications for the three AUC's.\label{tab3}}
\begin{tabular*}{\textwidth}{@{\extracolsep\fill}lllll@{}}
\toprule
\textbf{Area under the curve} & \textbf{Scenario 1}  & \textbf{Scenario 2}  & \textbf{Scenario 3}  & \textbf{Scenario 4}   \\
\midrule
Tumor Burden &
 0.979 &
0.998 & 
  0.062 & 
 0.006  \\
 \midrule
Survival &
 0.585 &
 0.020 & 
  0.587 & 
 0.020  \\
 \midrule
Total &
  0.986 &
 0.996 & 
  0.283 & 
 0.010  \\
\bottomrule
\end{tabular*}
\end{table*}
\end{center}

Similarly, in Scenario 2, where there is no treatment effect in the survival model, we would expect the rejection rate for the one-sided test under the null hypothesis of no treatment effect on survival AUC to be around $0.025$. As shown in Table \ref{tab3}, the rejection rate for testing the treatment effect on survival AUC in Scenario 2 is $0.020$, consistent with the corresponding rejection rate in Scenario 4. Likewise, in Scenario 1, where there is some treatment effect in both the longitudinal and survival models, we see that the rejection rates for the one-sided tests, as expected, are much higher than $0.025$.

Our parameter estimation results, based on $1000$ dataset replications and $Q = 25000$ iterations within each replication, are presented in Appendix \ref{AppendixA}. The bias and mean squared error (MSE) for parameter estimation are quite low. Among all parameters, the survival treatment effect coefficient, $\gamma_A$, consistently has the relatively larger MSE across all four simulation scenarios.

\section{Discussion}\label{sec7}
In this paper, we introduce a utility-based multi-component endpoint that integrates tumor burden biomarkers and survival data to estimate treatment effects in early-phase cancer clinical studies. The joint model for longitudinal tumor burden and survival data is partitioned into a survival model conditioned on baseline covariates and a longitudinal model conditioned on time-to-event data and baseline covariates. The key advantages of the proposed framework for treatment effect estimation are its intuitive mathematical interpretation of the composite utility-based endpoint in terms of the area-under-the-curve (AUC) and its ability to capture the average effects on both tumor burden and time-to-event outcomes, each measured in distinct units. Additionally, our Bayesian framework enhances computational efficiency, providing the posterior samples of treatment effects.

Our simulations demonstrated overall good performance of the proposed approach. In Scenario 4, where there is no treatment effect in either the longitudinal or survival models, the one-sided rejection rates were well below $0.025$ for all AUCs. Notably, we observed in Scenario 3 that if a treatment effect exists in the survival model only (with no treatment effect in the longitudinal model), there is some propagation of the effect from the survival model to the longitudinal model. This results from the way we decompose the joint distribution of longitudinal and survival data.

The models considered for the observed data distributions here are parametric, which may not capture the full complexity of real data. It would be straightforward to introduce assumption-lean flexible semi-parametric or non-parametric models for both longitudinal and survival data. For example, a Cox model could be applied to survival data, while mixture models or regression tree models could be explored for longitudinal data. Appealing to counterfactuals to give causal interpretations to treatment effects under suitable identification assumptions would be another extension of the current work. Finally, further development could focus on handling missing data, particularly informative dropout, and informative censoring.

\bibliographystyle{apalike}
\bibliography{ref}

\begin{thebibliography}{}

\bibitem[Betancourt, 2017]{betancourt2017conceptual}
Betancourt, M. (2017).
\newblock {A conceptual introduction to Hamiltonian Monte Carlo}.
\newblock {\em arXiv preprint arXiv:1701.02434}.

\bibitem[Carpenter et~al., 2017]{carpenter2017stan}
Carpenter, B., Gelman, A., Hoffman, M.~D., Lee, D., Goodrich, B., Betancourt, M., Brubaker, M.~A., Guo, J., Li, P., and Riddell, A. (2017).
\newblock {Stan: A probabilistic programming language}.
\newblock {\em Journal of statistical software}, 76.

\bibitem[Collett, 2023]{collett2023modelling}
Collett, D. (2023).
\newblock {\em {Modelling survival data in medical research}}.
\newblock Chapman and Hall/CRC.

\bibitem[Crowther et~al., 2013]{crowther2013joint}
Crowther, M.~J., Abrams, K.~R., and Lambert, P.~C. (2013).
\newblock {Joint modeling of longitudinal and survival data}.
\newblock {\em The Stata Journal}, 13(1):165--184.

\bibitem[Dejardin et~al., 2010]{dejardin2010joint}
Dejardin, D., Lesaffre, E., and Verbeke, G. (2010).
\newblock {Joint modeling of progression-free survival and death in advanced cancer clinical trials}.
\newblock {\em Statistics in Medicine}, 29(16):1724--1734.

\bibitem[Driscoll and Rixe, 2009]{driscoll2009overall}
Driscoll, J.~J. and Rixe, O. (2009).
\newblock {Overall survival: still the gold standard: why overall survival remains the definitive end point in cancer clinical trials}.
\newblock {\em The Cancer Journal}, 15(5):401--405.

\bibitem[Eisenhauer et~al., 2009]{eisenhauer2009new}
Eisenhauer, E.~A., Therasse, P., Bogaerts, J., Schwartz, L.~H., Sargent, D., Ford, R., Dancey, J., Arbuck, S., Gwyther, S., Mooney, M., et~al. (2009).
\newblock {New response evaluation criteria in solid tumours: revised RECIST guideline (version 1.1)}.
\newblock {\em European journal of cancer}, 45(2):228--247.

\bibitem[Group et~al., 2020]{ich2020e9}
Group, I. E.~W. et~al. (2020).
\newblock {ICH E9 (R1): addendum on estimands and sensitivity analysis in clinical trials to the guideline on statistical principles for clinical trials}.

\bibitem[Hogan and Laird, 1997]{hogan1997mixture}
Hogan, J.~W. and Laird, N.~M. (1997).
\newblock {Mixture models for the joint distribution of repeated measures and event times}.
\newblock {\em Statistics in medicine}, 16(3):239--257.

\bibitem[Imbens, 2004]{imbens2004nonparametric}
Imbens, G.~W. (2004).
\newblock {Nonparametric estimation of average treatment effects under exogeneity: A review}.
\newblock {\em Review of Economics and statistics}, 86(1):4--29.

\bibitem[Irwin, 1949]{irwin1949standard}
Irwin, J. (1949).
\newblock {The standard error of an estimate of expectation of life, with special reference to expectation of tumourless life in experiments with mice}.
\newblock {\em Epidemiology \& Infection}, 47(2):188--189.

\bibitem[Jiang et~al., 2024]{jiang2024bootstrap}
Jiang, L., Liu, X., Phillips, P.~C., and Zhang, Y. (2024).
\newblock {Bootstrap inference for quantile treatment effects in randomized experiments with matched pairs}.
\newblock {\em Review of Economics and Statistics}, 106(2):542--556.

\bibitem[Lawrence~Gould et~al., 2015]{lawrence2015joint}
Lawrence~Gould, A., Boye, M.~E., Crowther, M.~J., Ibrahim, J.~G., Quartey, G., Micallef, S., and Bois, F.~Y. (2015).
\newblock {Joint modeling of survival and longitudinal non-survival data: current methods and issues. Report of the DIA Bayesian joint modeling working group}.
\newblock {\em Statistics in medicine}, 34(14):2181--2195.

\bibitem[Neal, 2012]{neal2012mcmc}
Neal, R.~M. (2012).
\newblock {MCMC using Hamiltonian dynamics}.
\newblock {\em arXiv preprint arXiv:1206.1901}.

\bibitem[Otsu and Rai, 2017]{otsu2017bootstrap}
Otsu, T. and Rai, Y. (2017).
\newblock {Bootstrap inference of matching estimators for average treatment effects}.
\newblock {\em Journal of the American Statistical Association}, 112(520):1720--1732.

\bibitem[Rizopoulos, 2012]{rizopoulos2012joint}
Rizopoulos, D. (2012).
\newblock {\em {Joint models for longitudinal and time-to-event data: With applications in R}}.
\newblock CRC press.

\bibitem[Royston and Parmar, 2013]{royston2013restricted}
Royston, P. and Parmar, M.~K. (2013).
\newblock {Restricted mean survival time: an alternative to the hazard ratio for the design and analysis of randomized trials with a time-to-event outcome}.
\newblock {\em BMC medical research methodology}, 13:1--15.

\bibitem[Rubin, 1976]{rubin1976inference}
Rubin, D.~B. (1976).
\newblock {Inference and missing data}.
\newblock {\em Biometrika}, 63(3):581--592.

\bibitem[Rubin, 1981]{rubin1981bayesian}
Rubin, D.~B. (1981).
\newblock {The Bayesian bootstrap}.
\newblock {\em The annals of statistics}, pages 130--134.

\bibitem[Schoenfeld, 1983]{schoenfeld1983sample}
Schoenfeld, D.~A. (1983).
\newblock {Sample-size formula for the proportional-hazards regression model}.
\newblock {\em Biometrics}, pages 499--503.

\bibitem[Uno et~al., 2014]{uno2014moving}
Uno, H., Claggett, B., Tian, L., Inoue, E., Gallo, P., Miyata, T., Schrag, D., Takeuchi, M., Uyama, Y., Zhao, L., et~al. (2014).
\newblock {Moving beyond the hazard ratio in quantifying the between-group difference in survival analysis}.
\newblock {\em Journal of clinical Oncology}, 32(22):2380.

\bibitem[Wilson et~al., 2015]{wilson2015outcomes}
Wilson, M.~K., Karakasis, K., and Oza, A.~M. (2015).
\newblock {Outcomes and endpoints in trials of cancer treatment: the past, present, and future}.
\newblock {\em The Lancet Oncology}, 16(1):e32--e42.

\bibitem[Zhuang et~al., 2009]{zhuang2009overall}
Zhuang, S.~H., Xiu, L., and Elsayed, Y.~A. (2009).
\newblock {Overall survival: a gold standard in search of a surrogate: the value of progression-free survival and time to progression as end points of drug efficacy}.
\newblock {\em The Cancer Journal}, 15(5):395--400.

\end{thebibliography}


\newpage

\appendix

\section{Simulation results for model parameters}\label{AppendixA}

\begin{center}
\begin{table}[!h]
\caption{Parameter estimation results across 1000 simulation replications\label{table4}}
\begin{tabular}{llcccc}
\toprule
 & & \multicolumn{4}{c}{\bfseries \normalsize Simulation scenario 1}  \\
\cline{3-6}
SN & Parameter & True value & Estimation & Bias  & MSE  \\
\midrule
1 & $\sigma_{error}$           &    0.250   &     0.250  &   0.000  &  0.000     \\

2 & $\beta_{A}$        &   -2.250   &    -2.247  &   0.003  &  0.001      \\

3 & $\beta_{X}$         &   1.500   &     1.499  &  -0.001  &  0.000      \\

4 & $\gamma_{X}$         &    -1.200  &     -1.216  &  -0.016  &  0.000     \\

5 & $\gamma_{A}$         &    -0.750  &     -0.725  &   0.025 &   0.013     \\

6 & $\textbf{b}_{\mu1}$         &   -10.000   &   -10.013  &  -0.013  &  0.000      \\

7 & $\textbf{b}_{\mu2}$         &  11.000   &    11.017  &   0.017  &  0.000       \\

8 & $\textbf{b}_{sd1}$         &   1.000  &      0.999  &  -0.001  &  0.000      \\

9 & $\textbf{b}_{sd2}$         &    1.000   &     0.981  &  -0.019  &  0.000     \\
\bottomrule
\end{tabular}
\end{table}
\end{center}

\begin{center}
\begin{table}[!h]
\caption{Parameter estimation results across 1000 simulation replications\label{table5}}
\begin{tabular}{llcccc}
\toprule
 & & \multicolumn{4}{c}{\bfseries \normalsize Simulation scenario 2}  \\
\cline{3-6}
SN & Parameter & True value & Estimation & Bias  & MSE  \\
\midrule
1 & $\sigma_{error}$           &   0.250    &    0.252  &   0.002  &  0.000      \\

2 & $\beta_{A}$        &   -2.250  &     -2.255  &  -0.005  &  0.001      \\

3 & $\beta_{X}$         &    1.500  &      1.498  &  -0.002  &  0.000     \\

4 & $\gamma_{X}$         &   -1.200   &    -1.218  &  -0.018 &   0.001      \\

5 & $\gamma_{A}$         &    0.000   &    -0.036  &  -0.036  &  0.015     \\

6 & $\textbf{b}_{\mu1}$         &     -10.000  &    -10.027 &   -0.027  &  0.001    \\

7 & $\textbf{b}_{\mu2}$         &  11.000    &   11.042   &  0.042  &  0.002       \\

8 & $\textbf{b}_{sd1}$         &    1.000  &      0.994  &  -0.006  &  0.000     \\

9 & $\textbf{b}_{sd2}$         &     1.000   &     0.953  &  -0.047  &  0.002    \\
\bottomrule
\end{tabular}
\end{table}
\end{center}

\begin{center}
\begin{table}[!h]
\caption{Parameter estimation results across 1000 simulation replications\label{table6}}
\begin{tabular}{llcccc}
\toprule
 & & \multicolumn{4}{c}{\bfseries \normalsize Simulation scenario 3}  \\
\cline{3-6}
SN & Parameter & True value & Estimation & Bias  & MSE  \\
\midrule
1 & $\sigma_{error}$           &    0.250   &     0.250   &  0.000  &  0.000     \\

2 & $\beta_{A}$        &   0.000  &      0.003 &    0.003  &  0.001      \\

3 & $\beta_{X}$         &   1.500   &     1.499  &  -0.001  &  0.000      \\

4 & $\gamma_{X}$         &   -1.200   &    -1.216  &  -0.016  &  0.000      \\

5 & $\gamma_{A}$         &    -0.750    &   -0.725  &   0.025  &  0.013     \\

6 & $\textbf{b}_{\mu1}$         &   -10.000   &   -10.014  &  -0.014 &   0.000      \\

7 & $\textbf{b}_{\mu2}$         &    11.000   &    11.017   &  0.017  &  0.000     \\

8 & $\textbf{b}_{sd1}$         &   1.000    &    0.999  &  -0.001  &  0.000      \\

9 & $\textbf{b}_{sd2}$         &   1.000  &      0.980  &  -0.020  &  0.000      \\
\bottomrule
\end{tabular}
\end{table}
\end{center}

\begin{center}
\begin{table}[!h]
\caption{Parameter estimation results across 1000 simulation replications\label{table7}}
\begin{tabular}{llcccc}
\toprule
 & & \multicolumn{4}{c}{\bfseries \normalsize Simulation scenario 4}  \\
\cline{3-6}
SN & Parameter & True value & Estimation & Bias  & MSE  \\
\midrule
1 & $\sigma_{error}$           &    0.250   &     0.252  &   0.002  &  0.000     \\

2 & $\beta_{A}$        &  0.000   &    -0.005  &  -0.005 &   0.001       \\

3 & $\beta_{X}$         &    1.500    &    1.498  &  -0.002 &   0.000     \\

4 & $\gamma_{X}$         &    -1.200   &    -1.218 &   -0.018  &  0.001     \\

5 & $\gamma_{A}$         &    0.000   &    -0.035 &   -0.035  &  0.015     \\

6 & $\textbf{b}_{\mu1}$         &   -10.000  &    -10.028  &  -0.028  &  0.001      \\

7 & $\textbf{b}_{\mu2}$         &    11.000  &     11.042  &   0.042 &   0.002     \\

8 & $\textbf{b}_{sd1}$         &    1.000   &     0.995  &  -0.005 &   0.000     \\

9 & $\textbf{b}_{sd2}$         &   1.000   &     0.952  &  -0.048 &   0.002      \\
\bottomrule
\end{tabular}
\end{table}
\end{center}

\newpage
\section{Spaghetti plots and Kaplan-Meier (KM) curves} \label{AppendixB}

\begin{figure}[!h]
        \centering
        \includegraphics[scale=0.5]{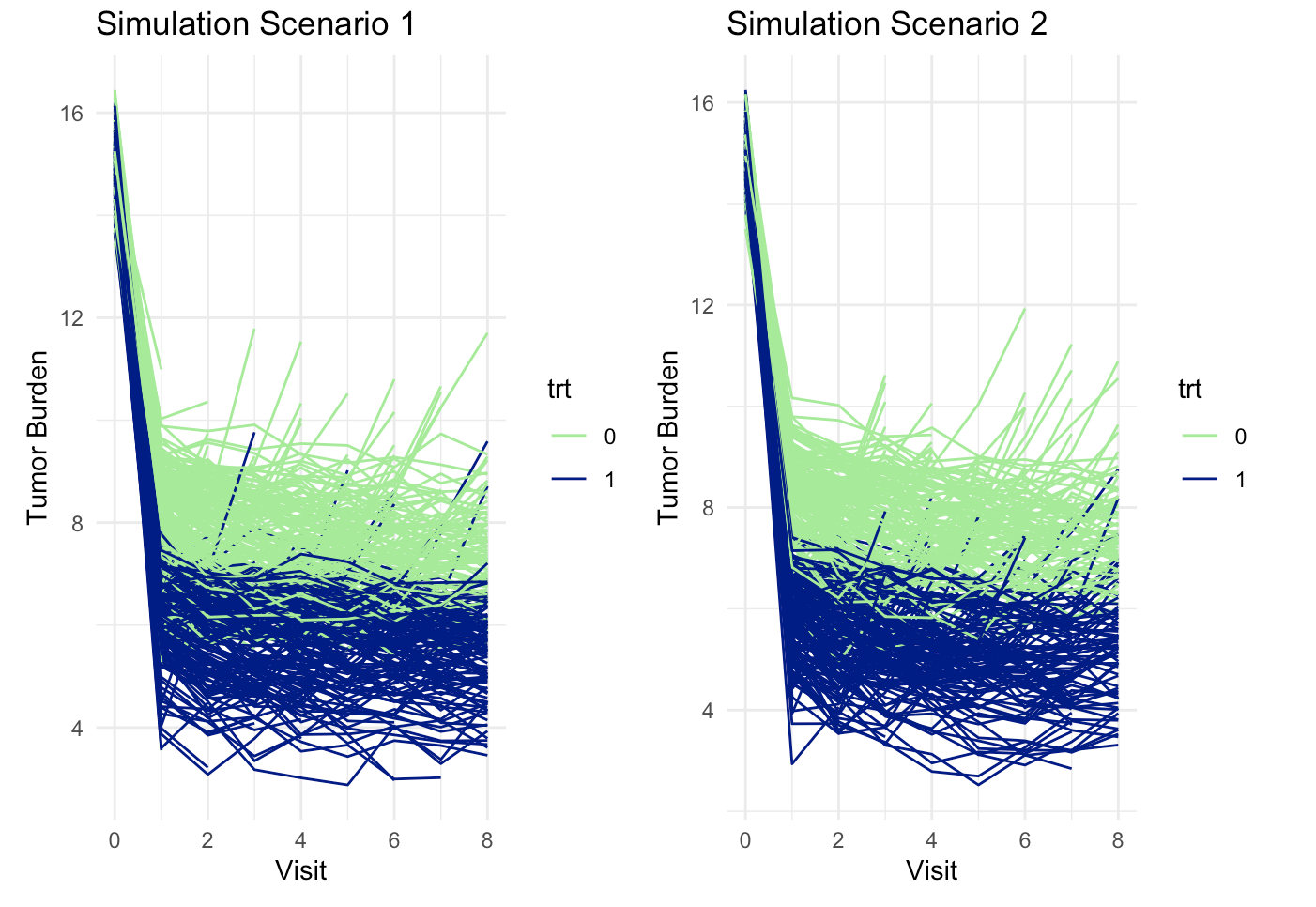}
        \caption{Spaghetti plots for tumor burden outcomes across simulation scenarios 1 and 2 from one data replication. These two scenarios assume some treatment effect on tumor burden.}
        \label{fig:TBcurvesWithBaseSim12}
    \end{figure}

   \begin{figure}[!h]
        \centering
        \includegraphics[scale=0.5]{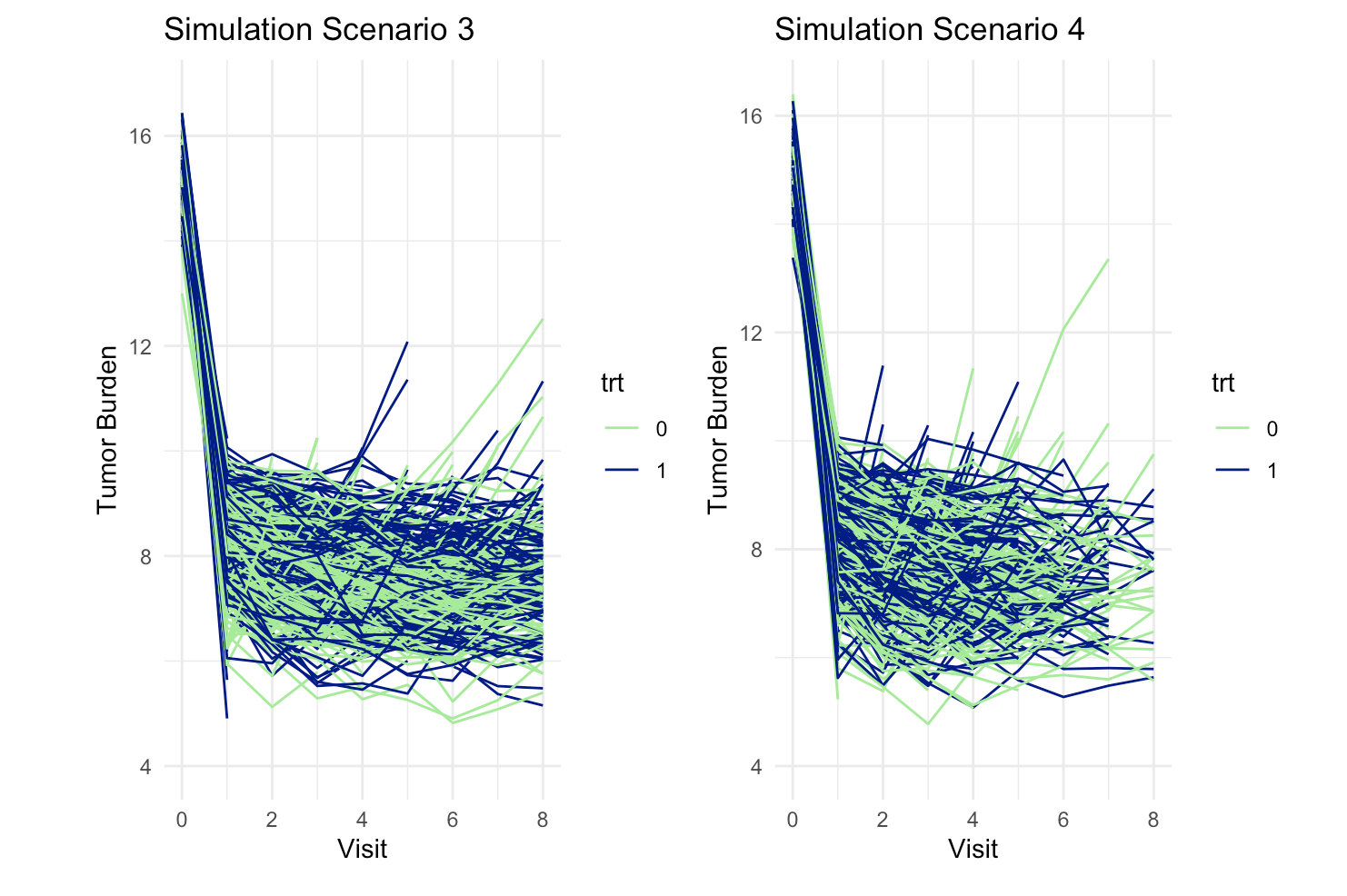}
        \caption{Spaghetti plots for tumor burden outcomes across simulation scenarios 3 and 4 from one data replication. These two scenarios assume no treatment effect on tumor burden.}
        \label{fig:TBcurvesWithBaseSim34}
    \end{figure} 

\begin{figure}[!h]
        \centering
        \includegraphics[scale=0.5]{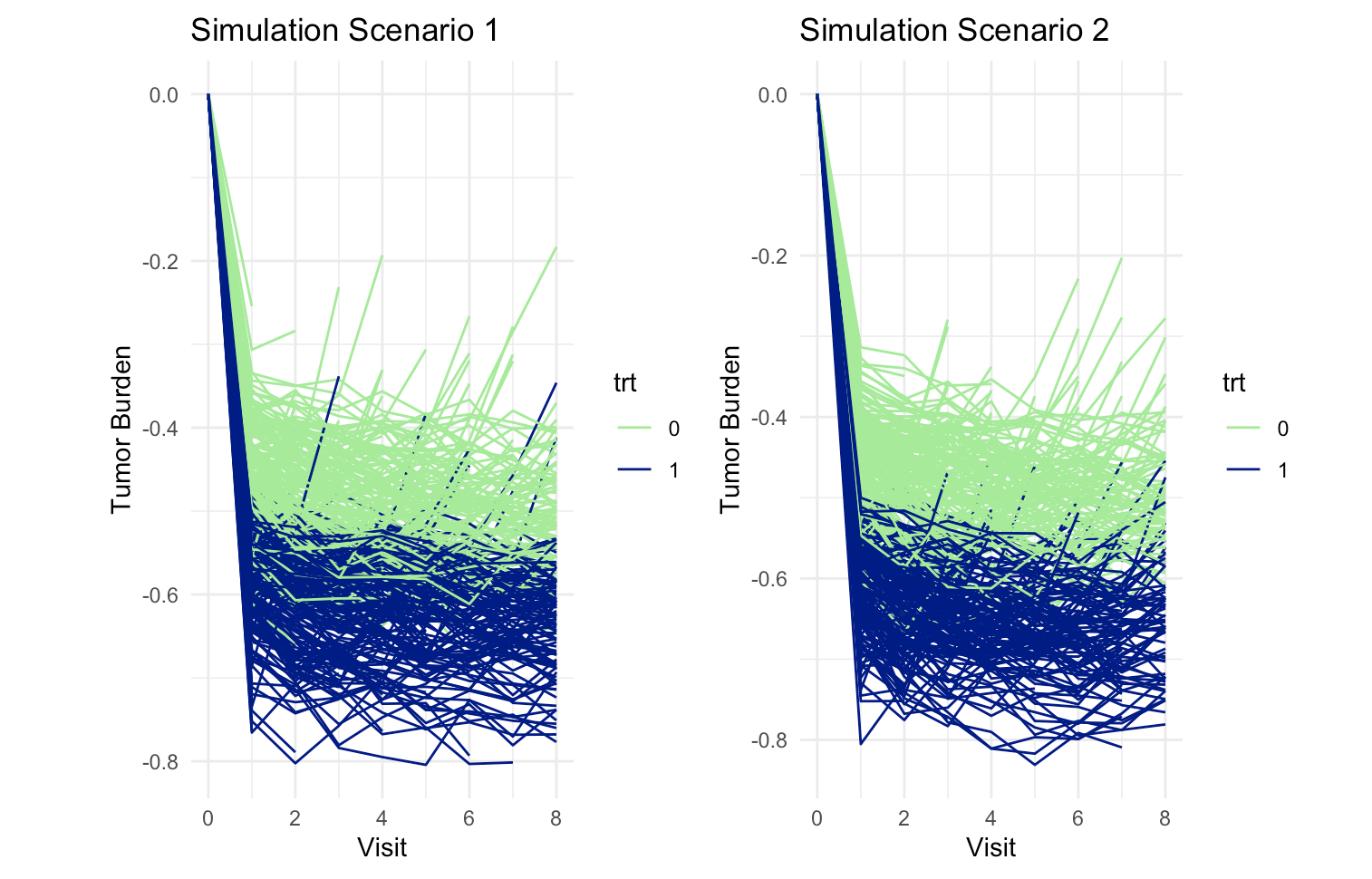}
        \caption{Spaghetti plots for tumor burden (change from baseline) outcomes across simulation scenarios 1 and 2 from one data replication. These two scenarios assume some treatment effect on tumor burden.}
        \label{fig:TBcurvesWithoutBaseSim12}
    \end{figure}

\begin{figure}[!h]
        \centering
        \includegraphics[scale=0.5]{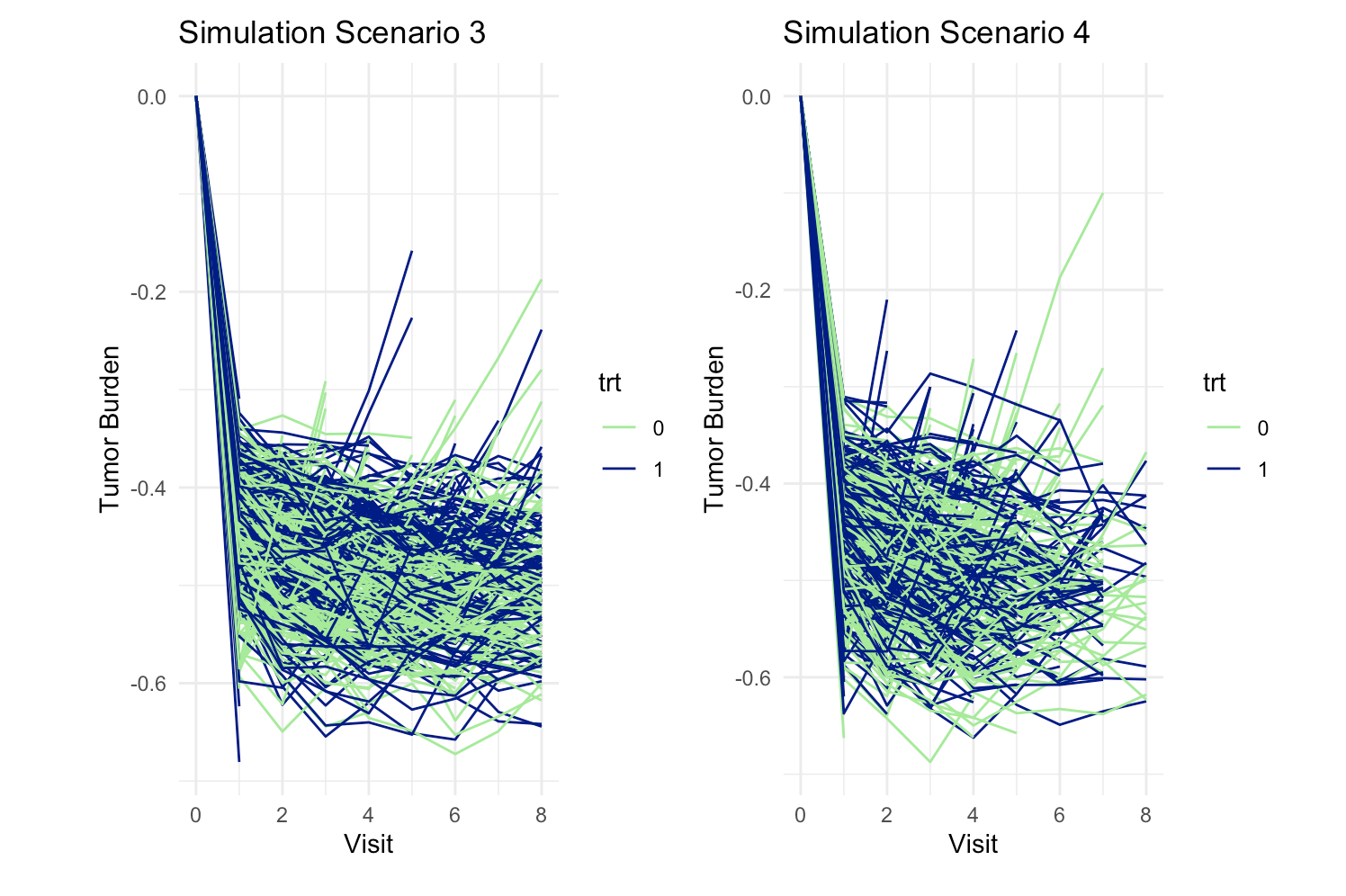}
        \caption{Spaghetti plots for tumor burden (change from baseline) outcomes across simulation scenarios 3 and 4 from one data replication. These two scenarios assume no treatment effect on tumor burden.}
        \label{fig:TBcurvesWithoutBaseSim34}
    \end{figure}

\begin{figure}[!h]
        \centering
        \includegraphics[scale=0.5]{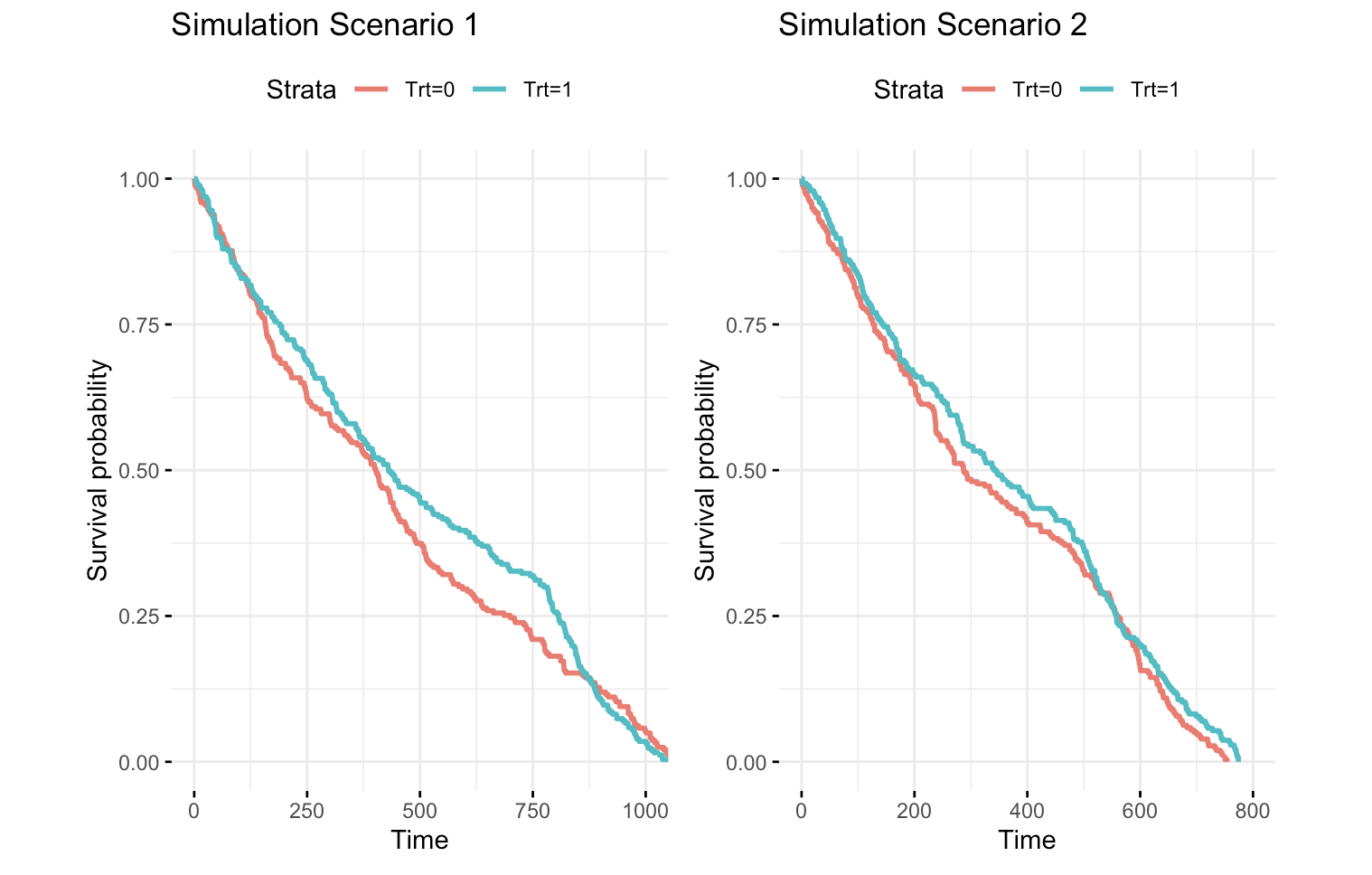}
        \caption{Kaplan-Meier (KM) curves for survival outcomes with administrative censoring across simulation scenarios 1 and 2 from one data replication. Scenario 2 assumes no treatment effect on the outcome.  The curves for two treatment groups may intersect as the proportional hazards assumption is not applied, which is common in clinical trials.}
        \label{fig:KMcurvesAdminCenSim12}
    \end{figure}

    \begin{figure}[!h]
        \centering
        \includegraphics[scale=0.5]{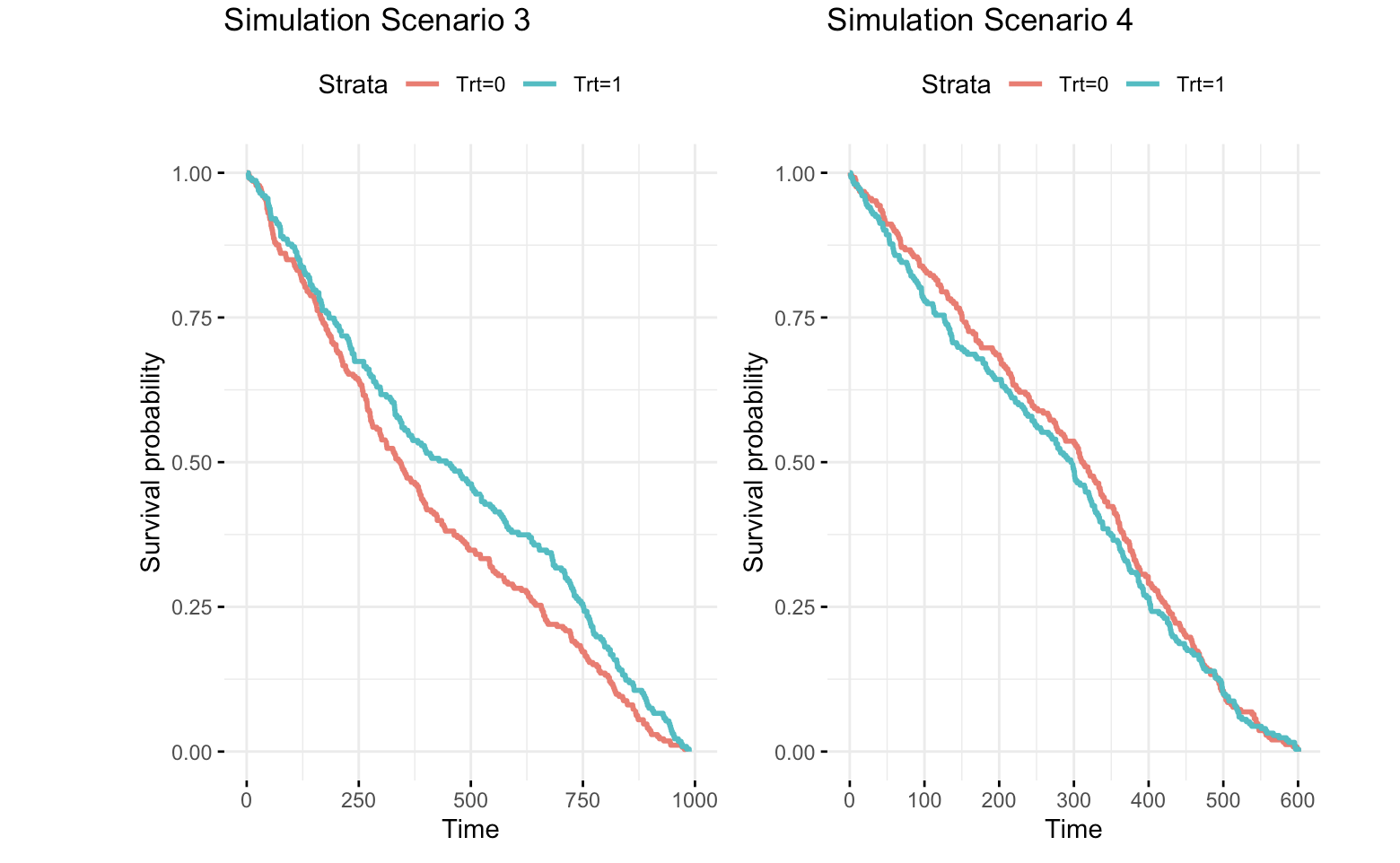}
        \caption{Kaplan-Meier (KM) curves for survival outcomes with administrative censoring across simulation scenarios 3 and 4 from one data replication. Scenario 4 assumes no treatment effect on the outcome.  The curves for two treatment groups may intersect as the proportional hazards assumption is not applied, which is common in clinical trials.}
        \label{fig:KMcurvesAdminCenSim34}
    \end{figure}

\begin{figure}[!h]
        \centering
        \includegraphics[scale=0.5]{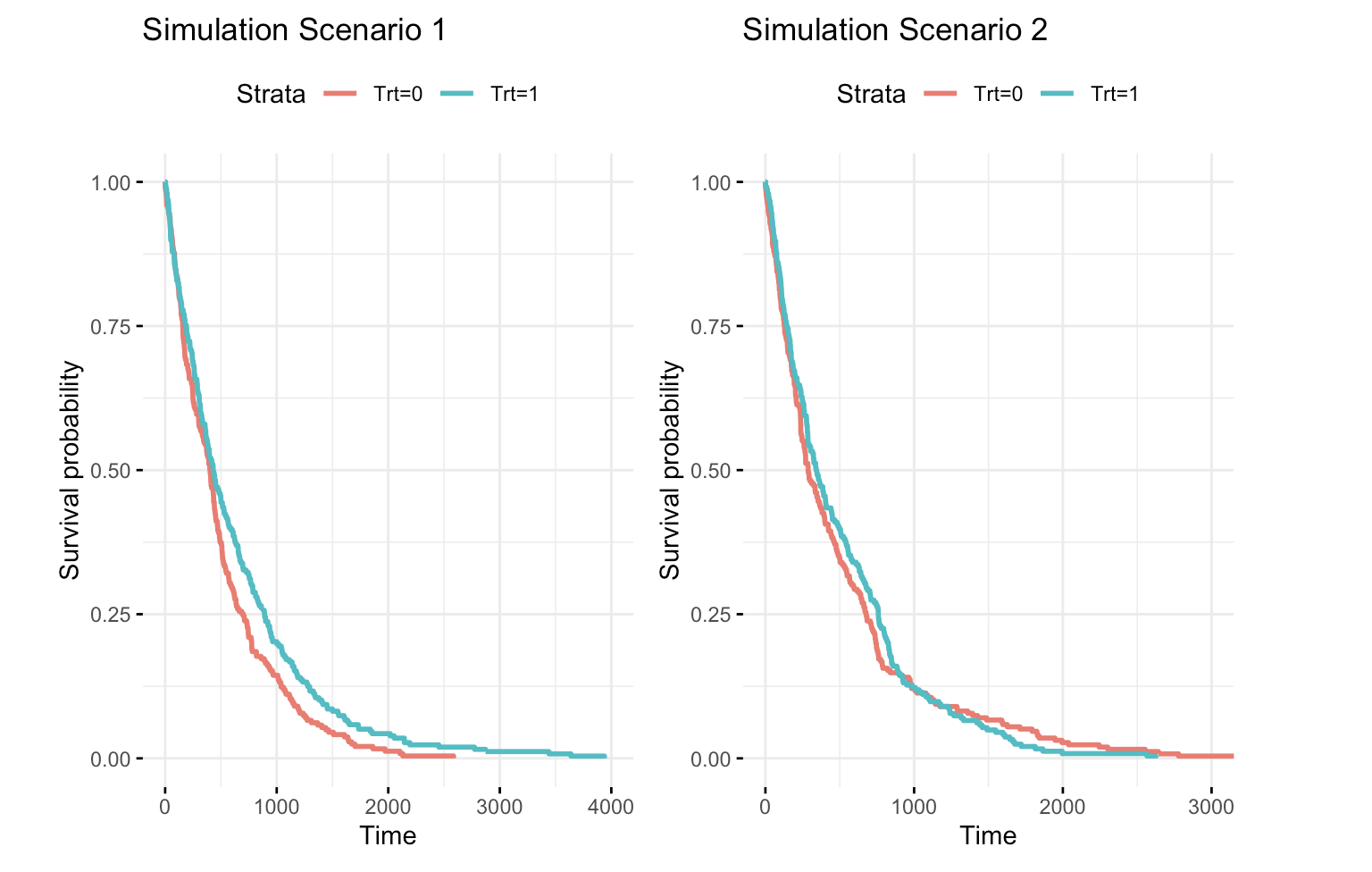}
        \caption{Kaplan-Meier (KM) curves for survival outcomes without administrative censoring across simulation scenarios 1 and 2 from one data replication. Scenario 2 assumes no treatment effect on the outcome.  The curves for two treatment groups may intersect as the proportional hazards assumption is not applied, which is common in clinical trials.}
        \label{fig:KMcurvesNoAdminCenSim12}
\end{figure}

\begin{figure}[!h]
        \centering
        \includegraphics[scale=0.5]{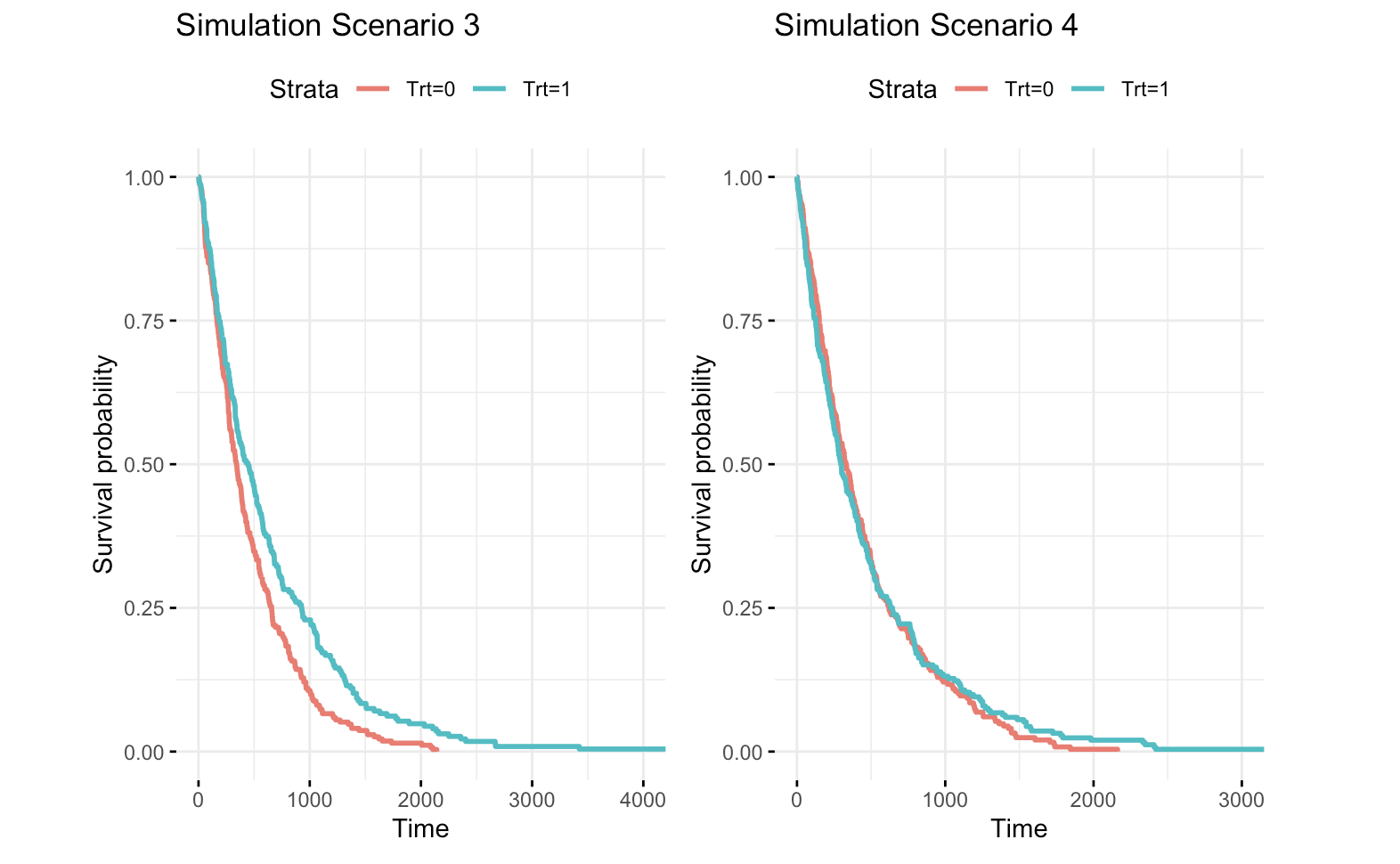}
        \caption{Kaplan-Meier (KM) curves for survival outcomes without administrative censoring across simulation scenarios 3 and 4 from one data replication. Scenario 4 assumes no treatment effect on the outcome.  The curves for two treatment groups may intersect as the proportional hazards assumption is not applied, which is common in clinical trials.}
        \label{fig:KMcurvesNoAdminCenSim34}
\end{figure}

\end{document}